# Crocs: Cross-Technology Clock Synchronization for WiFi and ZigBee

Zihao Yu, Chengkun Jiang, Yuan He, Xiaolong Zheng, Xiuzhen Guo
School of Software and TNLIST, Tsinghua University
{zh-yu17, jck15, guoxz16} @mails.tsinghua.edu.cn,
{heyuan, zhengxiaolong} @mail.tsinghua.edu.cn

## Abstract

Clock synchronization is a key function in embedded wireless systems and networks. This issue is equally important and more challenging in IoT systems nowadays, which often include heterogeneous wireless devices that follow different wireless standards. Conventional solutions to this problem employ gateway-based indirect synchronization, which suffers low accuracy. This paper for the first time studies the problem of cross-technology clock synchronization. Our proposal called Crocs synchronizes WiFi and ZigBee devices by direct cross-technology communication. Crocs decouples the synchronization signal from the transmission of a timestamp. By incorporating a barker-code based beacon for time alignment and cross-technology transmission of timestamps, Crocs achieves robust and accurate synchronization among WiFi and ZigBee devices, with the synchronization error lower than 1 millisecond. We further make attempts to implement different cross-technology communication methods in Crocs and provide insight findings with regard to the achievable accuracy and expected overhead.

## Categories and Subject Descriptors

C.2.2 [**Computer-Communication Networks**]: Network Protocols

## General Terms

Synchronization

## 1 Introduction

Time is critical information in embedded wireless systems and networks. Every network device maintains its own clock. Due to the inherent diversity of hardware and the difference among individual oscillators, the local clocks on different devices are essentially asynchronous. Clock synchronization, namely to synchronize different devices to a shared global clock, is therefore a key function in those networks. In event-driven application scenarios like environmental surveillance and safeguard, meaningful events will be falsely identified or even missed, if the sensor readings from different devices are not aligned in the temporal dimension. Clock synchronization is also crucial for correct operation of network protocols [1] and management [2].

There are various proposals on clock synchronization, depending on the structure, communication medium, and hardware of a network. NTP [21] is a representative protocol that is widely used in both wired and wireless networks. Clocks on the public NTP time servers are sent to the ordinary computers via layered multicast. Local filtering is required to resists faults and errors in the multi-casted clock data. RBS [5], TPSN [8], and FTSP [20] are protocols for clock synchronization in self-organized wireless networks, e.g. the wireless sensor network. The common idea is to share a global clock among the nodes. By periodically exchanging timestamps and accordingly calibrating their local clocks, the nodes can be synchronized with relatively high accuracy. Glossy [7] enhances the accuracy and efficiency of synchronization, by exploiting constructive interference of IEEE 802.15.4 symbols for fast network flooding. To summarize the existing proposal of clock synchronization, we can see a common methodology behind them: the synchronization process generally consists of two parts, namely transmission of timestamps and calibration of clocks. Direct communication among the devices is a precondition to make such a methodology effective.

The development of IoT applications make countless wireless devices deployed in the daily life space. Differing from the conventional wireless networks, IoT systems nowadays generally include heterogeneous devices that follow different wireless standards, e.g. WiFi, ZigBee, and BlueTooth. Clock synchronization is a crucial technique to enable collaboration and inter-operation among those devices. In an industrial IoT system, a pipeline with different manufacturing units and sensors must be controlled with accurate timing. Without clock synchronization of those units and sensors, the pipeline will be error-prone and inefficient, leading to undesired faults and losses.

Cross-technology clock synchronization (CTCS), namely to synchronizing the clocks of heterogeneous wireless de-vices, is a challenging problem. First of all, those devices following different wireless standards cannot communicate with each other directly. As a result, state-of-the-art designs rely on the shared gateway to forward timestamps. The uncertain end-to-end delay in transmitting timestamps via the gateway introduces additional synchronization error. Besides, the cost of such a gateway is usually over $100. In many scenarios where a gateway is not available, direct synchronization is desired.

Recent advances in cross-technology communication (CTC) shed light on the potential of direct data transmission across technologies. Time modulation [15] and energy modulation [10] have been proposed to modulate data from WiFi to ZigBee. Even with the ability of CTC, however, the conventional synchronization protocols cannot be applied in the context of CTCS. This is mainly due to the following reasons: First, CTC has relatively low throughput. The timestamp needs to be modulated by multiple wireless packets, which occupy the channel for a relatively long period of time. This potentially introduces additional but uncertain delays in local processing and network propagation. Second, CTC transmission is subjected to channel noise. Strong noise may corrupt the CTCS signals, leading to totally incorrect calibration of clocks. How to make a CTCS signal robust against noise is still an open problem.

In order to address the above challenges, in this paper we propose Crocs, a CTCS protocol for WiFi and ZigBee. To avoid the additional delay and the uncertainty in timestamp transmission, we decouple the synchronization signal from the transmission of a timestamp. The synchronization signal, which is realized as a sequence of WiFi beacons, precisely and reliably triggers an event of time alignment between WiFi and ZigBee. After that the timestamp of WiFi is transmitted to ZigBee by using CTC. The ZigBee device then executes clock calibration and calibrates its local clock (including parameters like *skew* and *offset*). The above synchronization process can be executed periodically so as to achieve a desired synchronization accuracy. The main contributions of this work are summarized as follows:

· To the best of our knowledge, Crocs is the first CTC-S protocol that works for WiFi and ZigBee. As we demonstrate in the evaluation, Crocs achieves robust and accurate synchronization, with the synchronization error lower than 1 millisecond. Crocs performs apparently better than the solutions currently applied in heterogeneous wireless networks.

· In Crocs, we propose a novel synchronization mechanism tailored to the CTCS scenario. Specifically, we design a barker-code based beacon to trigger the event of synchronization. The beacon is robust against channel noise and thus effective in ensuring the accuracy of time alignment between WiFi and ZigBee.

· We implement Crocs and carry out comprehensive evaluations. Besides validating the design of Crocs, our experiments provide insightful findings with regard to the achievable accuracy, when using different CTC methods for clock synchronization.

The rest of this paper is organized as follows. Section 2 discusses the related work. Section 3 presents the observational experiments on CTCS. We present an overview of the Crocs design in Section 4. Then we elaborate on the design of Crocs beacon for time alignment in Section 5 and two optional techniques for timestamp transmission in Section 6, followed by clock calibration in Section 7. Section 8 and Section 9 respectively present the analysis and evaluation results. We conclude this work in Section 10.

## 2 Related Work

Clock synchronization has been extensively studied in sensor networks [5, 6, 7, 8, 14, 20, 22]. RBS [5], TPSN [8] and FTSP [20] rely on disseminating the reference time to synchronize the whole network. Glossy [7] proposes radio-driven fast flooding to avoid the uncontrolled software delay and therefore improve the synchronization accuracy. Even though existing works have achieved high synchronization accuracy in sensor networks, they are infeasible for heterogeneous networks because the incompatibility hinders the reference timestamp dissemination across heterogeneous networks.

The CTC techniques are becoming popular today as they enable direct communication between heterogeneous devices following different wireless technologies without gateway [3, 4, 9, 28, 31, 32]. Based on the fact that ZigBee and WiFi devices coexist on the 2.4GHz band [24, 25], ZigBee nodes sense the radio signal strength (RSS) of on-air signals to identify the WiFi packet[12]. The existing studies propose time modulation or energy profile as the information carrier to exchange information [13, 29]. FreeBee [15] embeds symbols into beacons by shifting them from the reference time. GapSense [30] prepends customized preambles to the legacy packets. The receiver sense the gaps between energy pulses constructed by the sender and decode the me-ssage. Wizig [10] employs modulation techniques in both the amplitude and temporal dimensions to optimize the throughput over a noisy channel. Recent works such as WEBee [19] utilize the physical layer encoding to deliver message. We utilize CTC techniques to transmit the timestamp so that we are able to operate time synchronization between heterogeneous devices.

There have been studies about approaches to clock synchronization with the aid of external signal source [11, 16, 17, 18, 23, 26, 27]. The sensor nodes could sense signals from FM Radio Data System [16], fluorescent lighting [18] and power grid [23], extract useful information and then calibrate their own clocks. WizSync[11] makes the ZigBee sensors detect and synchronize to the periodical beacons broadcasted by Wi-Fi access points. In these works, the sensor nodes eliminate clock drift against other sensor nodes within the same network. Actually, the external signal sources only assist the synchronization of the sensor network. They are the facilitators rather than the participants, as the synchronization between heterogeneous devices is not involved.

We propose Crocs to synchronize the ZigBee and WiFi devices with both the offset and skew calibrated. In Crocs, the WiFi device works as time source rather than the external signal source. The ZigBee sensor nodes need to synchronize with the WiFi device. To the best of our knowledge, Crocs is

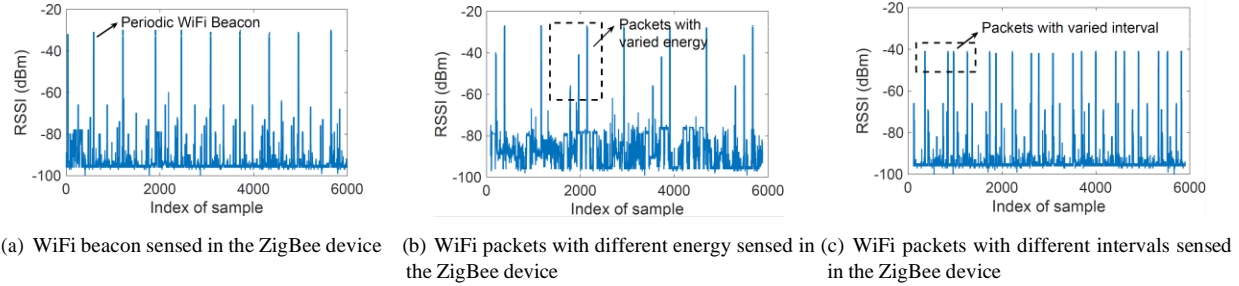

Figure 1. RSSI samples in the ZigBee device

(a) WiFi beacon sensed in the ZigBee device
(b) WiFi packets with different energy sensed in the ZigBee device
(c) WiFi packets with different intervals sensed in the ZigBee device

the first CTCS protocol that works for WiFi and ZigBee.

## 3 Observation

In this section, we present the analysis that motivates the design of the Crocs protocol and the empirical observations that serve as the cornerstone of Crocs.

### 3.1 Cross-Technology Sensing

The increasing demand of collecting synchronized data from heterogeneous data sources motivates us to study the cross-technology synchronization. We mainly consider the synchronization between the WiFi devices and ZigBee devices because of the ubiquity of these devices.

We first investigate how to deliver the synchronization timestamps for the heterogeneous devices. As we know, WiFi and ZigBee have totally incompatible PHY implementations, which makes it impossible for them to directly decode each other's packets. Fortunately, recent studies on CTC shed the light of enabling direct information exchanges among heterogeneous devices. WiFi and ZigBee coexist on the 2.4GHz ISM band and their channels overlap with each other on the frequency band. Therefore, while WiFi devices are transmitting packets, ZigBee nodes can sense high RSS in the channel. Moreover, when WiFi devices adapt the interval or the transmission power of transmitting packets, the RSS sensed by ZigBee will change accordingly. And we know that the RSSI sampling is a foundational MAC function equipped in most wireless standards including WiFi and ZigBee.

To illustrate the ability of sensing the RSS change, we set up a USRP device to send WiFi packets with a fixed interval and use the RSSI sampling function in a ZigBee device to record the RSSI sequence in Figure 1. In the figure, we can find the periodic high RSSI samples are caused by the transmission of the WiFi packets. In Figure 1(a), we could find the periodic high RSSI is caused by the transmission of the WiFi beacon. The WiFi beacon is sent at a fixed beacon interval. Furthermore, the RSS sensing ability of the ZigBee devices can also capture the energy change in the WiFi packets in Figure 1(b). We set three levels of WiFi transmission power and the ZigBee devices can clearly distinguish them, which makes it possible for WiFi device to modulate information by adapting the transmission power. We further adjust the transmission interval of WiFi packets and the ZigBee devices can still sense the interval change in Figure 1(c). Thus the WiFi can also embed symbols into the beacon interval as the ZigBee can decode the symbols by the interval of high RSS. In conclusion, the rich sensing ability of the ZigBee devices provides the opportunities for WiFi devices to communicate with them in an indirect way.

### 3.2 Cross-Technology sensing under noise

Actually, even if we can realize CTC, the throughput of the communication is very low. In order to transmit a synchronization timestamp, many WiFi packets are required, which obstacles the traditional clock synchronization mechanism that encapsulates the timestamp into a single packet as shown in Figure 2. Because the long cross-technology transmission makes it difficult for the receiver and sender to record the timestamp at the same time. Therefore, the traditional synchronization mechanism is infeasible in CTCS. To realize CTCS, a light-weight and accurate synchronization mechanism must be designed to minimize uncertainty of timestamp recording.

To avoid the uncertainty caused by the packets used for timestamp transmission, we decouple the synchronization process. We first use a mutually detectable event to notify both devices of a synchronization time point and then transmit the timestamp. Thus the timestamp is recorded at the mutually detectable event rather than the start of the timestamp transmission. However, it is challenging to design a unique event for both devices to be recognized. Despite the rich sensing ability of the ZigBee devices as in Figure 1, in reality, there are lots of WiFi devices sending packets over the air so the RSS sampling in our daily environment could be different. Because of the interference and noise in the environment, it is quite challenging to design a unique and reliable enough synchronization message so that the ZigBee device could accurately identify the synchronization time point.

**In summary**, the low throughput of the CTC methods makes it infeasible to directly use traditional synchronization mechanism in the CTC scenarios. Instead, it is required that a decoupled synchronization mechanism in which we first use a unique WiFi packet to notify both devices to record its local time and then we send the timestamp of one device to another one with CTC methods. However, the unique packet can be easily broken by random WiFi packets in the environment. So a reliable and efficient mechanism should be designed to realize the notification of the time alignment.

## 4 Design Overview

Based on the observation above, we design Crocs, which enables the ZigBee device to synchronize to the WiFi device. In Figure 3, we show the work flow of Crocs. In Crocs, the WiFi device initiates the synchronization and goes into Time Alignment phase. The WiFi device first generates beacon packets to create a mutually detectable pattern and records the start time of the beacon packets. ZigBee devices then identify the customized beacon and calculate the start time of receiving the first beacon packet. Then the WiFi and ZigBee devices agree on the unique global time point by recording the time of the identical event (the first packet of our customized beacon). We call the pair of timestamps recorded on both sides as a clock synchronization pair.

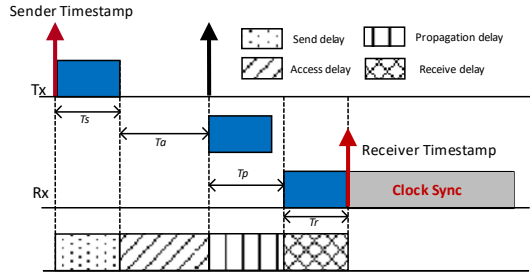

**Figure 2. Traditional clock synchronization for homogeneous devices**

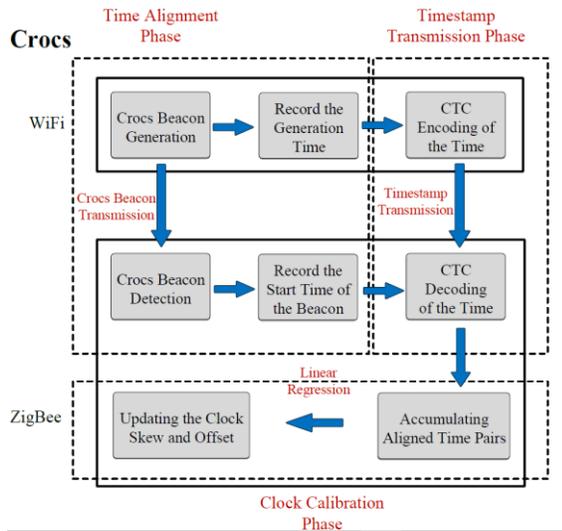

**Figure 3. The design overview of our system**

After agreeing on the time of the identical event, the WiFi device goes into Timestamp transmission phase and transmits its recorded timestamp to the ZigBee device for further synchronization. To deliver the timestamp timely and reliably, we explore the temporal modulation and energy modulation. We make special designs of timestamp transmission so that Crocs is tailored to cross-technology synchronization.

Upon receiving WiFi's timestamp, the ZigBee device enters the clock calibration phase and adjusts the global time. By cumulating several synchronization clock pairs, the ZigBee device is able to synchronize its clock to WiFi device locally before next synchronization to reduce the accumulated error. We refer to the time interval between the synchronization clock pairs as pair interval.

## 5 Crocs Beacon

As explained in Section 3, it is inaccurate to use traditional synchronization mechanism in which devices transmit a single packet containing the timestamp for cross-technology synchronization. In this section, we explain our design of the customized Crocs beacon as an identical alignment event to make WiFi and ZigBee devices obtain the alignment timestamps with minimum error.

The design of Crocs beacon is challenging in three aspects. (1) The communication standards of WiFi and ZigBee are incompatible. A mutually sensible beacon that can be detected by the heterogenous devices should be designed. (2) The general environments where WiFi and ZigBee devices operate are noisy due to the coexisting devices. Then the sensible beacon should be robust and distinguishable even under noisy environments. (3) The beacon must be succinct and can be detected in real time to avoid uncontrollable time error.

It is well known that a ZigBee device cannot directly decode the WiFi packets and is only able to sense the energy change caused by these packets. A straightforward method is to exploit the high energy of WiFi packets as alignment event and then ZigBee and WiFi devices record the start or end time of the high energy as the time alignment point. However, on 2.4GHz band, many different wireless standards like WiFi, ZigBee, and Bluetooth coexist. The transmissions of any coexisting device can also cause high energy on channel and causes the false detection of alignment event on ZigBee devices. So it is unreliable to use just a single high energy packet. Furthermore, even if we can know the high energy is caused by the WiFi packets, the ZigBee device is still unable to decide which WiFi device it wants to synchronize with because it is common that multiple WiFi devices sending the beacons. The false detected alignment events will corrupt the clock synchronization pairs (defined in Section 4) and lead to the failure of cross-technology synchronization.

In order to guarantee the ZigBee device to accurately align with the designated WiFi device, we propose to construct a special sequence of WiFi packets as the Crocs beacon. The candidate parameters of WiFi devices that we can manipulate are transmission power and packet interval. A packet interval denotes the time interval between two WiFi packet transmissions. The first method is to transmit the packets in the beacon with different transmission powers so that the ZigBee receiver can observe the variations of received signal strength indicator (RSSI) during the beacon. However, we find that using the energy of the packets to create the beacon pattern is unreliable because the packet RSSI is easily affected by the channel noise and interference. Such unstable RSSI sequences cannot meet the requirement of reliability.

The second method is to transmit the packets with coded packet intervals to create the pattern of interval sequences. Detecting the existence of packets is much more reliable than distinguishing different RSS levels, as we only need one energy threshold of RSSI to identify packet existence, and several thresholds to distinguish different RSS levels. Therefore, we investigate creating alignment beacon by constructing specific packet interval sequence.

**Table 1. Packet intervals for different beacon length**

| Beacon Length | Packet Intervals |
|---|---|
| 3 | $(t_1, t_2)$ |
| 4 | $(t_1, t_1, t_2)$ |
| 5 | $(t_1, t_1, t_1, t_2)$ or $(t_1, t_1, t_2, t_1)$ |
| 6 | $(t_1, t_1, t_1, t_2, t_1)$ |
| 8 | $(t_1, t_1, t_1, t_2, t_2, t_1, t_2)$ |
| 12 | $(t_1, t_1, t_1, t_2, t_2, t_2, t_1, t_2, t_2, t_1, t_2)$ |
| 14 | $(t_1, t_1, t_1, t_1, t_1, t_2, t_2, t_1, t_1, t_2, t_1, t_2, t_1)$ |

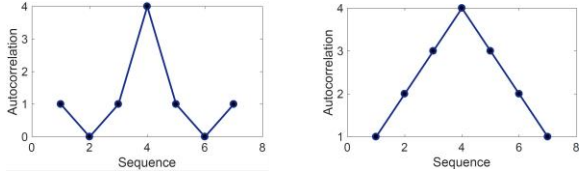

(a) Autocorrelation for the Barker code $(+1, +1, +1, -1)$  (b) Autocorrelation for the sequence $(+1, +1, +1, -1)$

**Figure 4. Autocorrelation for different sequences**

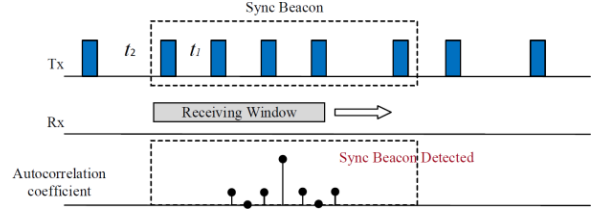

**Figure 5. 5-packets beacon detection at the ZigBee device**

We select two atomic packet intervals $t_1$ and $t_2$ that are different with the regular WiFi packet interval. By selecting the intervals that seldom exists in regular WiFi traffic, the beacon has certain reliability. To further improve the reliability of the beacon, we use the two atomic packet intervals to define a set of beacons (Crocs beacon) with various beacon lengths, as shown in Table 1. Here, the beacon length refers to the number of WiFi packets in one Crocs beacon. In Table 1, we define the sequence of packet intervals according to the Barker sequence if we regard $t_1$, $t_2$ as $+1, -1$ correspondingly. The Barker sequence are known for its ideal autocorrelation property. For all $1 \leq v < N$, the autocorrelation coefficients

$$c_v = \sum_{j=1}^{N-v} b_j b_{j+v}$$

are smaller than 1. $b_j$ if $j = 1, 2, \ldots, N$ is the Barker sequence. Take a Barker sequence $(+1, +1, +1, -1)$ with length 4 for example, the autocorrelation reaches its peak when $j = 4$ and all the other autocorrelations remain smaller than 1 as shown in Figure 4(a). Except Barker sequence, other sequence with length 4 does not have the same property. We demonstrate the autocorrelation for $(+1, -1, +1, -1)$ in Figure 4(b). We could use the property of the Barker code to detect our interested beacons and avoid the interference of other sequences.

Using Barker code can help the ZigBee receiver to accurately find the start of the Crocs beacon and to record the accurate alignment time point. On the ZigBee device side, the ZigBee nodes continue sensing the channel and collect the RSSI sequences. It maintains a parameter to record the average RSSI for normal WiFi packet. When the ZigBee node detects continuous samples of an obviously higher RSSI, it will record it as a beacon packet. As shown in Figure 5, after cumulating enough beacon packets, the ZigBee node calculates the autocorrelation of the RSSI sequence during a moving window, based on the above predefined beacons. When the sequences match the predefined packet sequences of a beacon, the autocorrelation will present its peak value at the start of the beacon sequence.

## 6 Transmission of Timestamps

After the time is aligned by the start of the identical beacon, the WiFi devices will transmit its timestamp to the ZigBee receiver. As analyzed in Section 3, to deliver the timestamp timely without gateways, direct CTC from WiFi to ZigBee is necessary. In this section, we introduce the temporal modulation and energy modulation tailored to cross-technology synchronization. Note that we do not use energy modulation for Crocs beacon as the beacon is used to align WiFi and ZigBee to a unique time point. Sending a beacon via temporal modulation (Barker code) guarantees this property, while energy modulation cannot.

### 6.1 Temporal Modulation

In this section, we introduce the timestamp transmission based on temporal modulation. In temporal modulation, WiFi device modulates the information by controlling the transmission time of packets. The ZigBee device senses the RSS in the channel and identifies the RSSI of WiFi packets. It demodulates the timestamp by analyzing the time features of the RSS sequences.

*6.1.1 Modulation Method*

In Crocs, the time on WiFi devices is the typical NTP time. A NTP timestamp is a 64-bit symbol sequence. We embed the symbols into the timing of packets by controlling the interval of the packets. The WiFi device sends a sequence of packets and adjusts the interval between packets according to the symbols to transmit. The ZigBee device senses the channel, collects the RSSI sequence and then demodulates the symbol embedded in the intervals between the peaks of signal strength.

To improve the transmission efficiency, Crocs tries to keep the encoding period as short as possible. However, the USRP device we use to send packets has a minimum packet interval of 10 milliseconds. So we have to ensure that the packet interval is larger than 10 milliseconds regardless of the symbol we send in Crocs. Moreover, the scheme we propose should be robust to the noise in the channel.

Considering these requirements, we propose the transmission method based on temporal modulation. Notice that the timestamp we want to send can be regarded as a series of numbers. To transmit this value, we can divide it by digits and then encode the digits sequentially. We modulate every digit by controlling the interval between the packets. Specifically, if the digit we want to send is $a$ $(0 < a < 9)$, we set the packet interval to $(10 + a)$ milliseconds and the next packet interval to $(20 - a)$ milliseconds. Thus we use two intervals to modulate one digit and ensure that the modulation window of one digit is fixed for 30 milliseconds.

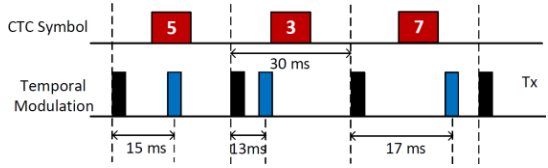

**Figure 6. Example temporal modulation**

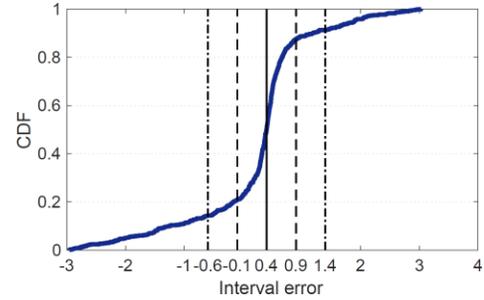

**Figure 7. The CDF of packet interval jitters**

Figure 6 presents a simple example of our modulation. To encode one digit, we actually control the transmission time of the second packet in one window. By identifying the position of the second packet, we can decode the transmitted bit. The digits modulated in Figure 6 are 5, 3 and 7. From the figure, we can also find our temporal modulation has two features. First, our temporal modulation can use different packet intervals according to the hardwares. In this example, the packet interval is longer than 10 ms due to the restrictions of USRP we use. If the hardware supports shorter packet interval, we can adjust the parameter to use a short packet interval.

Second, the time it takes to encode one digit is fixed to 30 ms. Such a window length makes our method more robust as the practical packet interval is not stable due to the sender delay of the USRP device. When the ZigBee device detects that the duration of one symbol differs greatly from 30 milliseconds, it will abandon this symbol as this modulation suffers from the abrupt sender delay.

*6.1.2 The Adjustment of Encoding Time Granularity*

As stated above, we actually encode the digit at the granularity of 1 ms. If the time jitter is larger than 0.5 milliseconds, then the symbol we transmit may be incorrectly decoded as another symbol. For example, when we transmit the digit 3 using our encoding method, the sender sets the packet interval as 13 milliseconds and 17 milliseconds and transmits three packets continuously. If the middle packet suffers from a sudden large sender delay, the interval of RSS peak detected by the receiver can have large errors, e.g. 13.7 ms and 16.3 ms during one of our experiments. Then the receiver will incorrectly decode this symbol as 4 rather than 3.

To avoid this situation and improve robustness, a straightforward method is to increase the encoding time granularity to tolerate larger time jitters. We observe the practical packet interval jitters to determine whether it is necessary and how large to adjust the encoding time granularity.

We set the WiFi device to send packets with a fixed period of 15 ms, and then measure the practical interval of high RSS caused by the packets which the ZigBee device detects. We observe the distribution of the jitters of packet intervals in practice, compared with the predetermined period (15 ms).

Figure 7 shows the experiment results. We can find that the detected intervals are usually larger than the predetermined intervals. We also find that the practical intervals appear to obey a normal distribution with non-zero mean. Thus to guarantee that the symbol can be demodulated accurately, we propose two methods to reduce the time error. The first method is to compensate for the detected packet intervals. In the scenario presented in Figure 7, as the time error concentrates on the range around 0.4 ms, we set the compensation value as 0.4 ms. So if we subtract 0.4 ms from the detected intervals, the time error between them and preset period will fit the normal distribution with the mean value of zero. The second method is to increase the granularity of modulation as mentioned before. In our original scheme, the encoding period is 30 ms with the encoding time granularity of one millisecond. To improve the robustness, we increase the encoding time granularity to 2ms and double all the parameters in our scheme. More concretely, to transmit a number $a(0 < a < 9)$, we send three packets with the first packet interval set as $20 + 2a$ ms and second as $40 - 2a$ ms. Then the modulation window of one digit is 60 ms, twice of the original window. In return, the acceptable error range of time also double from 0.5ms to 1ms.

To adopt larger granularity improves the fault tolerance by sacrificing the throughput as the encoding period will be longer. Since the time jitter is caused by software delay that is device-dependent, we can adjust the parameters of temporal modulation for different devices under different environments. We will further study performance improvement with the time granularity adjustment in the evaluation.

**6.2 Energy Modulation**

The major drawback of the temporal modulation method is the low throughput caused by its long time window of encoding one bit. An alternative method of timestamp transmission is energy modulation. The WiFi sender controls the packet transmission power to create different energy levels on the channel. We encode the packet absence as bit 0 and packet presence as bit 1 when two energy levels are used. When using multiple energy levels, the sender can encode multiple bits into different energy levels simultaneously. The coding efficiency of energy modulation is higher than the efficiency of temporal modulation. Therefore, the throughput of energy modulation will be improved.

Figure 8 presents a simple example of energy modulation with two energy levels. The encoded CTC symbols are 1011. Then the WiFi sender generate a packet transmission schedule to create the packet presence in the first, third and fourth time slots and the packet absence in the second time slot. The time slot here refers to the time window of every bit. The ZigBee receiver keeps sampling the channel and collect the RSSI samples. Due to the high energy caused by packet transmission, the ZigBee receiver will detect high RSS on channel during the first, third and fourth time slots and the low energy in the second time slot. Then the corresponding encoded symbols 1011 can be decoded correctly.

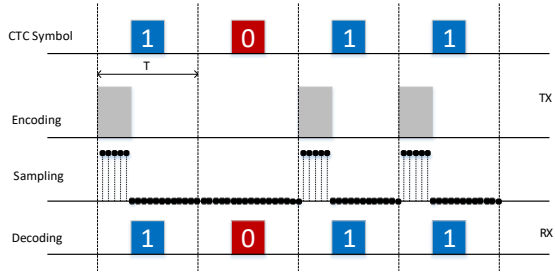

**Figure 8. Example of energy modulation**

The energy modulation achieves high accuracy in environments under low noise. We will further evaluate this in the Section 9. Besides the high-throughput energy modulation, Crocs also integrates incremental clock transmission that only transmits the difference between current time and the last timestamp. When the synchronization interval is less than one minute, the time difference won't exceed the scope of $2^{26}$. Then Crocs can only transmit 26 bits to represent the time difference. The size of transmitted timestamp will therefore decrease from 64 bits to 26 bits except the initial timestamp transmission. The transmission time can be further reduced.

## 7 Clock Calibration

Last but not least, the clock calibration component is designed to reduce the time error with a timer model.

We use $T_w$ and $T_z$ to denote the clock of WiFi and ZigBee respectively. Then the relationship between $T_w$ and $T_z$ is:

$$T_z = T_w + t_{offset} \quad (1)$$

The $t_{offset}$ here denotes the clock offset between the clocks. Notice that the clocks of WiFi device and ZigBee device do not have the exact same frequency, so the clock offsets between them are not constant. Suppose that the clock of ZigBee runs faster than WiFi, then the offset between ZigBee and WiFi will grow with time. Fortunately, the offset between the two clocks changes in a linear fashion provided the short term stability of the clocks is good [8]. For two synchronization clock pairs $(T_{z1}, T_{w1})$ and $(T_{z2}, T_{w2})$, we have the equations below:

$$T_{z1} = T_{w1} + t_{offset1}$$

$$T_{z2} = T_{w2} + t_{offset2}$$

$$t_{offset2} = t_{offset1} + t_{skew} * (T_{z2} - T_{z1})$$

Here we consider the clock of WiFi to be the global clock and the ZigBee device needs to calibrate its clock against it. The reason will be illustrated in Section 8. We denote the clock skew between WiFi and ZigBee by $t_{skew}$. From the above equations, the ZigBee device is able to calculate its clock offset and skew against the WiFi device using the clock synchronization pairs. Therefore, clock calibration component leverages multiple clock synchronization pairs to build the timer model and estimate the clock offset and skew. In Crocs, we use linear regression on the past several data points to build the timer model.

## 8 Analysis
### 8.1 Feasibility and Applicability of Crocs

In Crocs, we enable the clock of the ZigBee device to synchronize with the clock of a WiFi device. However, the reverse synchronization is not included. Actually, we are able to realize the reverse synchronization with the same design if we substitute the ZigBee packet for the WiFi packet. Our design is based on two observations. First, in practical scenarios, the power supply of WiFi devices is more stable than that of ZigBee devices that are always equipped with a few batteries. Moreover, the oscillators in WiFi devices are more stable than those in ZigBee devices because of the higher price. Therefore, the clock of the WiFi devices is usually more stable and accurate. The second observation is that WiFi devices are more prevalent than ZigBee devices and WiFi devices can directly connect with the Internet. The ultimate goal of the IoT is to make everything connect with the Internet, so if we synchronize the ZigBee device with the WiFi device, we could establish a unified clock for all the devices in the Internet. In consideration of this, we design Crocs to synchronize the ZigBee nodes with the WiFi nodes. Crocs may co-work with the synchronization protocols in respective networks (e.g. NTP [21], FTSP [8]). The combination improves synchronization accuracy and flexibility to synchronize front-end heterogeneous devices.

## 9 Evaluation
### 9.1 Implementation

We implement a prototype of Crocs on a USRP platform and TelosB, a commercial ZigBee platform. We use an USRP N210/GNURadio to generate WiFi packets following IEEE 802.11 standards. In our implementation, we choose 802.11 channel 6 and 802.15.4 channel 17 to produce the spectrum overlap. We use one USRP/N210 sender to send the packets to operate cross-technology clock synchronization. To explore the performance under noise, we use another USRP/N210 to generate Gaussian noise with different power. The RSSI sampling rate of TelosB node is 6 KHz. Notice that we do not need any modifications of hardware to make Crocs work on a commercial device.

### 9.2 Crocs beacon Evaluation

We evaluate the reliability of Crocs beacon by measuring the beacon matching rate under different noise levels. The USRP device sends the beacon to the ZigBee device under different noise levels. Every time we send 2000 beacons to calculate the matching rate. In the experiment, three beacon lengths 3, 4, and 5 are used to measure the matching rate. As for the packet interval, we fix the $t_1$ to 30 ms and set $t_2$ to 50 ms, 60 ms and 70 ms. The experimental results are demonstrated in Figure 9. From the results, we observe that with the beacon length fixed, the bigger the gap between $t_1$ and $t_2$ is, the higher matching rate we will achieve. The reason is that the uncertainty of the time delay in the MAC layer

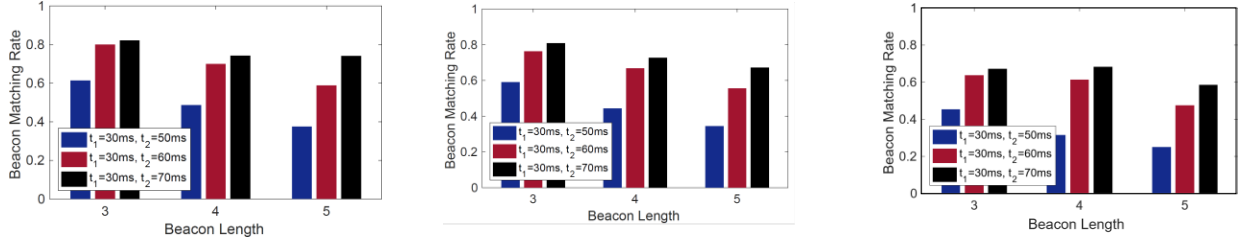

(a) The beacon matching rate with low-level noise  (b) The beacon matching rate with medium-level noise  (c) The beacon matching rate with high-level noise at sender will increase the actual packet interval. So the $t_1$ interval could be falsely detected as $t_2$ interval. Increasing the gap between $t_1$ and $t_2$ could lower the detection error. However, larger gap of the intervals will force the ZigBee devices to sample longer time as shown in Table 5, which is time-consuming for the resource-limited TelosB nodes.

**Figure 9. The beacon matching rate with different parameters**

We also observe from Figure 9 that the matching rate decreases as we increase the beacon length. To explain this, we look into the experimental data and find that the packet intervals from a USRP sometimes deviate from the predefined values by up to ten milliseconds. We guess that USRP suffers from limited memory and computational resources so that the timing control of the packet interval could abruptly change when transmitting packets. The beacon with longer length is more likely to encounter the interval changes, which lowers the matching rate.

Generally, we observe that the beacon length of 3 with $t_2$ to be 70 ms achieves the best matching rates, which are 0.8219, 0.8082, 0.6712 when the noise level is low, medium and high respectively.

### 9.3 Timestamp Transmission Evaluation

In this section, we utilize two modulation techniques to transmit the timestamp and explore the performance of them. We use the bit error rate (BER) to evaluate the reliability of the timestamp transmission.

*9.3.1 Temporal Modulation Evaluation*

To make the ZigBee device demodulate the timestamp accurately, we utilize the two methods proposed in the Section 6 to reduce the BER. We conduct three sets of experiments to measure the BER of timestamp transmission with time granularity of 1ms, 2ms and 3ms respectively. The results are shown in Figure 10. As Figure 10 shows, the decoding accuracy is rather low when we adopt 1 ms as the encoding granularity. The BER apparently decreases when we double the granularity. And to set the encoding granularity as 3 produces the BER of about 0.0022.

*9.3.2 Energy Modulation Evaluation*

In this section we evaluate the BER of energy modulation to transmit the timestamps. Figure 11 presents the BER with different energy levels and slot length. The BER of energy modulation is lower when we adopt only two energy levels rather than four. Also, if we set longer slot length, we will get relatively lower BER. The BER of temporal modulation with the encoding time granularity of 3ms is close to that of energy modulation with the energy levels of 2. The BERs of temporal modulation and energy modulation are both lower than 0.01 with appropriate parameters. Therefore, both the modulation methods are reliable to transmit the timestamps from WiFi devices to ZigBee devices.

### 9.4 Clock Calibration Evaluation

In this section we evaluate the performance of the clock calibration by the time error after one synchronization. We compare our design with the simple clock synchronization, in which the ZigBee device only uses one synchronization clock pair to calibrate its clock offset. The performance of simple clock synchronization is showed in Figure 12, The time error grows to nearly 200 ms after 7 s and exceeds 1 s only after 44 s. This would mandate continuous re-synchronization with a period of less than one second to keep the error below 30 ms, which is a significant overhead in terms of bandwidth and energy consumption.

In our clock calibration design, the ZigBee device collects 5 timestamps every time to conduct linear regression and then updates both the skew and offset. From the results shown in Figure 13, we find that the time error is well controlled within 1 ms. Compared to the results of simple clock synchronization, the time error does not grow over time. We conclude that the skew between ZigBee device and WiFi device has been accurately compensated for the controlled time error. Therefore, the measured skew value is very close to the ground truth and the remaining error is mainly caused by the non-determined delay mentioned in Section 8. Moreover, as the time error keeps below 1 ms after 43 s before next synchronization, the ZigBee node sense the channel for 0.5s and sleep throughout the rest of a synchronization period (at least 43s). The overhead is less than 1.25% in duty cycle. The clock calibration significantly reduce the synchronization frequency, which helps the resource-limited ZigBee nodes to save the synchronization cost.

### 9.5 Overall Evaluation

In this section, we evaluate the performance of the whole Crocs protocol with the metric of time error between the ZigBee device and the WiFi device. As we are the first to synchronize the ZigBee device to the WiFi device, we cannot compare our protocol with related works. So we observe the performance of Crocs with different timestamp interval defined in Section 4. We set the timestamp interval as 50 milliseconds and 7 seconds respectively, and then measure the time error every 7 seconds. We re-synchronize the two devices every time we measure the time error. The results

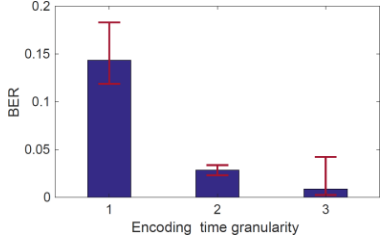
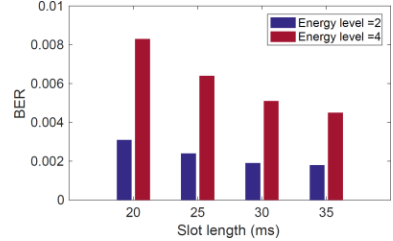
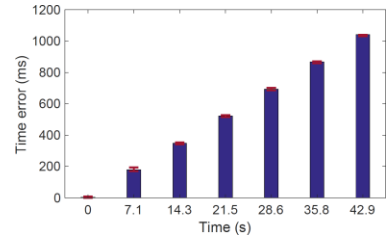

Figure 10. The BER of temporal modulation with different parameters

Figure 11. The BER of energy modulation with different parameters

Figure 12. The time error between two devices quickly grow over time without skew calibration

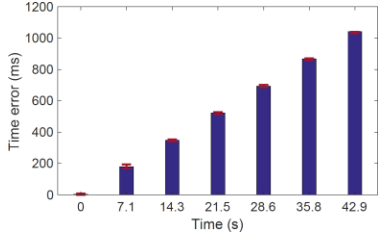
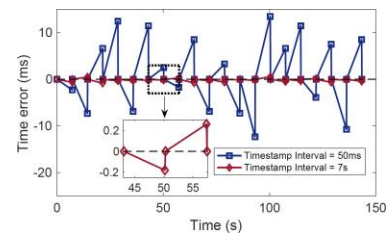
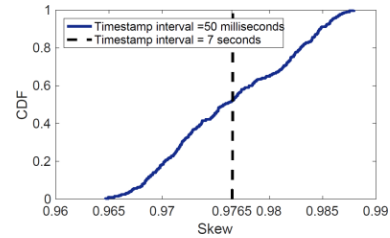

Figure 13. The time error stays stable with skew calibration

Figure 14. The time error under different timestamp Interval

Figure 15. The skew we measure under different timestamp Interval

are presented in Figure 14.

From Figure 14, we find that if we set the timestamp period as seven seconds, the time error is stable below one millisecond. But if we set the timestamp to 50 milliseconds, the time error will quickly grow with time after calibration. We even discover that the time error may grow above 10 milliseconds before the next synchronization. To explain this, we further analyze the results and find that the performance depends on the skew calibration. The skew reflects the frequency difference between two devices. We define the difference between the skew that we calculate and the ground truth as the skew error. If the skew error is large, then the time error will grow quickly as the frequency difference is not well compensated. Figure 15 shows the distribution of skew with the timestamp period of 50 milliseconds and 7 seconds respectively. When we set the timestamp period to 7 seconds, the clock skew is relatively stable and concentrate on the range between 0.9765 and 0.9766. Judging from the synchronization performance in Figure 14, the ground truth of skew between the two devices lies in the range. So when we set the timestamp to 7 milliseconds, we successfully calibrate the skew and offset of the ZigBee device against WiFi device, because the time error stays below one millisecond all the time. But when we set the timestamp to 50 milliseconds, the skew deviates from the ground truth so the frequency difference leads to the increasing time error until the next synchronization.

## 10 Conclusion

We propose Crocs, a synchronization protocol that enables the clock synchronization between heterogeneous devices. We design the Crocs beacon so that heterogeneous devices could record the timestamps at the alignment time point in complex environment and then leverage cross-technology communication to transmit timestamps. Base on this, the receiver cumulates the timestamps and synchronizes to the sender by using linear regression. We also implement a prototype of Crocs on a software radio platform and a commercial ZigBee device. The evaluation shows that Crocs achieves the synchronization accuracy of lower than 1 ms. By calibrating the clock skew, we can lower the synchronization frequency to reduce the synchronization cost in the resource-limited ZigBee nodes.

## 11 Acknowledgements

This work was supported by National Key R&D Program of China 2017YFB1003000, National Basic Research Program (973 program) under Grant of 2014CB347800, National Natural Science Fund of China for Excellent Young Scientist No.61422207, The NSFC No. 61772306, No. 61472382, No. 61672320, and China Postdoctoral Science Foundation No. 2016M601034.